\documentclass[preprint,review,authoryear,12pt]{elsarticle}
 
\usepackage{url,lineno,algorithm,graphicx,stfloats,fixltx2e,eqparbox,mdwtab,amsmath,array,algorithmic,amssymb,subfig}

\journal{Planetary and Space Science}

\begin{document}

\begin{frontmatter}




\title{On the unmixing of MEx/OMEGA hyperspectral data}


\author[ekpa,noa]{Konstantinos E. Themelis}
\ead{themelis@space.noa.gr}
\address[ekpa]{Department of Informatics and Telecommunications, University of Athens,  Ilissia, 157 84 Athens, Greece}
\address[noa]{Institute for Space Applications and Remote Sensing, National Observatory of Athens, 152 36, P. Penteli, Greece}
\author[paris1,paris2]{Fr\'{e}d\'{e}ric Schmidt}\ead{frederic.schmidt@u-psud.fr}
\author[noa]{Olga Sykioti} \ead{sykioti@space.noa.gr}
\author[noa]{Athanasios A. Rontogiannis} \ead{tronto@space.noa.gr}
\author[noa]{Konstantinos D. Koutroumbas} \ead{koutroum@space.noa.gr} 
\author[noa]{Ioannis A. Daglis} \ead{daglis@space.noa.gr}
\address[paris1]{Univ. Paris-Sud, Laboratoire IDES, UMR 8148, Orsay, F-91405, France}
\address[paris2]{CNRS, Orsay, F-91405, France}

\begin{abstract}
This article presents a comparative study of three different types of estimators used for supervised linear unmixing of two MEx/OMEGA hyperspectral cubes. The algorithms take into account the constraints of the abundance fractions, in order to get physically interpretable results. Abundance maps show that the Bayesian maximum a posteriori probability (MAP) estimator proposed in \cite{2008_themelis} outperforms the other two schemes, offering a compromise between complexity and estimation performance. Thus, the MAP estimator is a candidate algorithm to perform ice and minerals detection on large hyperspectral datasets.
\end{abstract}


\end{frontmatter}


\section{Introduction}
\label{sec:intro}
The surface of Mars is currently being imaged with a combination of high spectral and spatial resolution. This gives the ability to detect and map chemical components on the Martian surface and atmosphere more accurately than before. Spectral unmixing (SU) is one of the techniques currently used for this purpose, \cite{2002_keshava, 2008_moussaoui, 2010_schmidt}. SU is the procedure by which the measured spectrum of a mixed pixel is decomposed into a number of constituent spectra, called endmembers, and the corresponding fractions, or abundances, that indicate the proportion of the presence of each endmember in the pixel, \cite{2002_keshava}. Linear SU, which adopts the hypothesis that the spectrum of a mixed pixel is a linear combination of its endmembers' spectra, is more commonly used in practice. Based on a physical interpretation, two hard constraints are imposed on the abundance fractions of the materials in a pixel; they should be non-negative and sum to one. 

Several SU techniques for the unmixing of OMEGA (Observatoire pour la Min\'{e}ralogie, l' Eau, les Glaces et l' Activit\'{e} hyperspectral images), \cite{2004_bibring}, have been recently proposed in the bibliography. These techniques can be categorized into unsupervised, where a special procedure is first executed to get the endmembers' spectral signatures from the image, and supervised, where a priori knowledge of the image endmembers is available. A recent example of an unsupervised technique is the Bayesian source separation method, developed in \cite{2010_schmidt}. This technique is based on a Gibbs sampling scheme to perform Bayesian inference. Due to its high computationally complexity, a special implementation strategy is developed for its application to a complete OMEGA image data set, \cite{2010_schmidt}. Band ratio is the most commonly used supervised technique to detect minerals (\cite{2004_bibring,2004_bibring1,2005_langevin}). However some multiple endmember linear spectral unmixing algorithms have been proposed such as MELSUM, \cite{2008_combe}.
MELSUM uses a reference library containing various spectral signatures of minerals (used as endmembers), and is based on the classical spectral mixture analysis (SMA) algorithm. In \cite{2007_kanner}, a modified Gaussian model (MGM) has been exploited to estimate the fractional abundances of a compositionally diverse suite of pyroxene spectra in the martian surface. A wavelet based method has also been applied for the unmixing of hyperspectral martian data in \cite{2006_gendrin,2007_schmidt}.

In this paper, we focus on the problem of supervised SU. Our main objective is to estimate the abundances of the endmembers that are present in two OMEGA images, subject to the non-negativity and sum-to-one constraints. In the following, three different supervised unmixing algorithms are considered, namely the ENVI-SVD method, \cite{1989_boardman}, a quadratic programming (QP) technique, \cite{1996_coleman}, and a recently proposed Bayesian maximum a posteriori probability soft-constraint (MAPs) estimator, \cite{2008_themelis}. These algorithms are applied on two different hyperspectral OMEGA data sets, and they are evaluated through their corresponding abundance maps. The endmember reference spectra used for supervised unmixing are either extracted from the image itself, or selected from a spectral library of pure minerals. The experimental results show that the MAPs estimator results in abundance values that satisfy the constraints of the problem and provides a compromise between the performance of the QP technique and the complexity of the ENVI-SVD method. An earlier version of this paper was presented at the 2010 European Planetary Science Congress, \cite{2010_themelis}.


\section{Linear spectral unmixing techniques}
\label{sec:tech}
Before we present the unmixing techniques considered in this paper, a short description is presented here on the linear mixing model (LMM), \cite{2002_keshava}, which is assumed in all three of them. In a hyperspectral image, each pixel is represented by a $L$- dimensional vector ${\mathbf y}$, where $L$ is the number of the available spectral bands. The elements of ${\mathbf y}$ correspond to the reflectance measured at the respective spectral bands. The LMM assumes that the received pixel's spectrum is generated by a linear combination of endmembers' spectra. Suppose that the spectral signatures of $p$ materials that may exist in the image are  available. Then, ${\mathbf y}$ can be expressed by the following linear regression model:
\begin{equation}
\label{eq:model}
{\mathbf y} = {\mathbf \Phi} {\mathbf x} + {\mathbf n},
\end{equation}
where ${\mathbf \Phi} = \left[ \boldsymbol \phi _{1} \boldsymbol \phi_{2} \ldots \boldsymbol \phi_{p}\right]\in\mathbb{R}^{L\times p}_{+}$, is the mixing matrix containing the endmembers' spectra ($L$-dimensional vectors $\boldsymbol \phi_i, i = 1, 2, \dots, p$), ${\mathbf x}$ is a $p\times 1$ vector with the corresponding abundance fractions, and $ {\mathbf n}$ is a $L\times 1$ additive noise vector. 

Adopting the linear model in (\ref{eq:model}), three different unmixing algorithms are applied to the OMEGA data sets: i) a singular value decomposition method (ENVI-SVD), \cite{1989_boardman}, available in the ENVI image processing software ii) a QP technique, \cite{1996_coleman}, available in the Matlab environment, and iii) a recently proposed Bayesian MAPs estimator, \cite{2008_themelis}. These algorithms are briefly described in the following subsections.


\subsection{ENVI-SVD}
\label{sec:envi}
ENVI-SVD is a constrained least squares approach to the unmixing problem. Using the singular value decomposition (SVD) algorithm, the pseudo-inverse of the mixing matrix $\boldsymbol\Phi$ is computed. Then, the abundance fractions are easily estimated by multiplying the pseudo-inverse matrix with each image pixel's spectral vector. The advantage of this method is its low computational complexity, since the pseudo-inverse matrix is computed only once as a preprocessing step, and is then applied to all image pixel vectors. As far as the sum-to-one constraint is concerned, it is imposed to the problem using an extra (weighted) equation to the linear system of equations (\ref{eq:model}). However, ENVI-SVD does not take into account the non-negativity of the abundances, which can result in  negative abundance values that have no physical meaning.

\subsection{Quadratic programming technique}
\label{sec:qp}
This quadratic programming technique is a reflective Newton method, which minimizes the quadratic function of the least squares error of the unimixing problem, subject to the sum-to-one and non-negativity constraints imposed on the abundances. The QP technique is based on an iterative optimization scheme, which has to be repeated separately for each image pixel vector. Keeping in mind that a hyperspectral image may be composed of thousands of pixels, solving a separate optimization problem for each pixel adds up to the computational complexity of the method. An implementation of the quadratic programming technique is available in the optimization toolbox of Matlab.

\subsection{MAPs estimator}
\label{sec:maps}
The MAPs estimator proposed in \cite{2008_themelis} is a Bayesian estimator specifically designed to address the inverse problem of supervised hyperspectral unmixing. In a Bayesian framework, appropriate prior distributions are assigned to the unknown parameters of the estimation problem, which usually reflect the parameters' natural characteristics. The Bayesian approach of \cite{2007_benavoli} is adopted in which a Gaussian distribution is used as a prior for the abundance vector $\mathbf{x}$ and the MAP estimator is then utilized. By exploiting the symmetry of the problem's convex constraints, the parameters of the Gaussian posterior distribution (i.e., the mean and the covariance matrix) can be expressed in closed forms, \cite{2008_themelis}. Due to the statistical nature of the Bayesian estimator, the constraints are not explicitly imposed to the estimated parameters. To alleviate this, the final step of the algorithm is a projection of the MAP estimation point on the polytope of constraints, providing its nearest estimate that satisfies the constraints, \cite{2008_themelis}. This algorithm has substantially lower complexity than the QP technique, since it relies on the computation of simple closed-form expressions.  

\section{Discussion}
\label{sec:results}

The previously described algorithms are applied to two different OMEGA data cubes: (a) a scene of Mars' South Polar Cap, and (b) a Syrtis Major observation. The OMEGA instrument is a spectrometer on board ESA's Mars Express satellite, which provides hyperspectral images of the Mars surface, with a spatial resolution from 300m to 4km, 96 wavelength channels in the visible band and 256 wavelength channels in the near infrared band, \cite{2004_bibring}. OMEGA uses three different detectors, with spectral resolutions about $7.5$nm in the $0.35 - 1.05\mu$m wavelength range (visible and near infrared channel or VNIR), $14$nm between $0.94$ and $2.70\mu$m (short wave infrared channel or SWIR) and an average of $21$nm from $2.65$ to $5.2\mu$m (long wave infrared channel or LWIR), respectively. The two hyperspectral data sets, the reference spectra used for each image and the SU results obtained from the application of the three methods are analytically described in the following sections.


\subsection{South Polar Cap image cube}
\label{sec:south_cap}

\begin{figure}[ht]
\centering
\includegraphics[width=\linewidth]{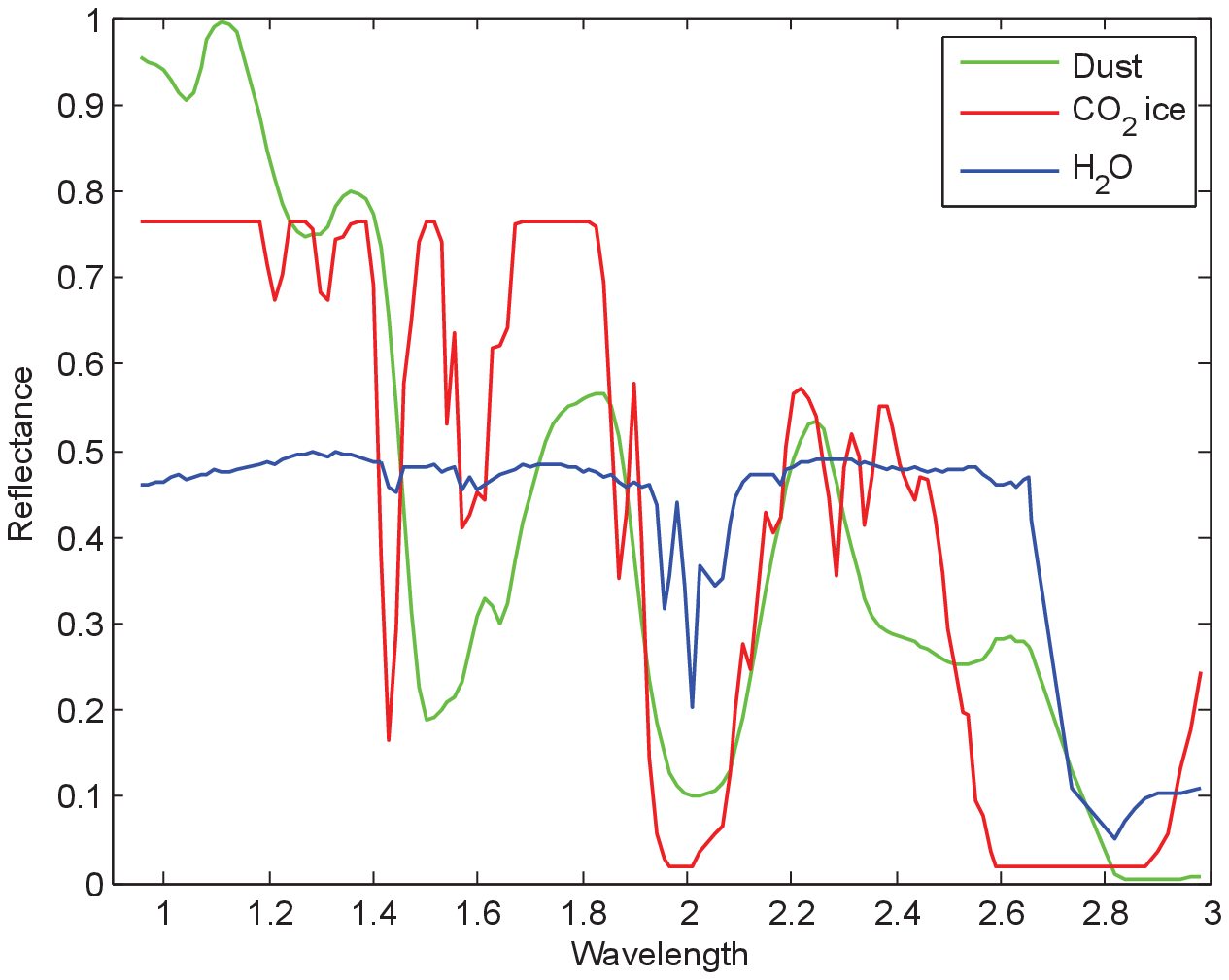}
\caption{Reference spectra of the South Polar Cap OMEGA image. The available endmembers are: (a) OMEGA typical dust materials with atmosphere absorption. (b) synthetic CO2 ice with grain size of 100 mm, (c) synthetic H2O ice with grain size of 100 microns.}
\label{fig:endmbr_spectra01}
\end{figure}

This data set consists of a single hyperspectral data cube obtained by looking towards the South Polar Cap of Mars in the local summer (Jan. 2004). The data cube is made up of two channels: 128 spectral planes from $0.93$ to $2.73 \mu$m with a resolution of $14$nm and $128$ spectral planes from $2.55$ to $5.11 \mu$m with a resolution of 21nm. Noisy bands were excluded, and 156 out of the 250 initial bands were finally utilized in the region from $0.93$ to $2.98\mu$m to avoid the thermal emission spectral range. The linear model mixing matrix consists of the following three reference spectra: (a) CO$_2$ ice (synthetic data with grain size = $100$mm), (b) H$_2$O ice (synthetic data with grain size = $10\mu$m), and (c) dust, which were all detected a priori using the Wavanglet method of \cite{2007_schmidt}. These endmembers are discussed in the first OMEGA publication, \cite{2004_bibring1}, and are also verified by \cite{2010_schmidt}, using the Bayesian positive source separation method of \cite{2008_moussaoui}. The respective spectral signatures of the endmembers are shown in Fig. \ref{fig:endmbr_spectra01}. 



\begin{figure}[t]
\centering
\subfloat[]{\includegraphics[height=160mm]{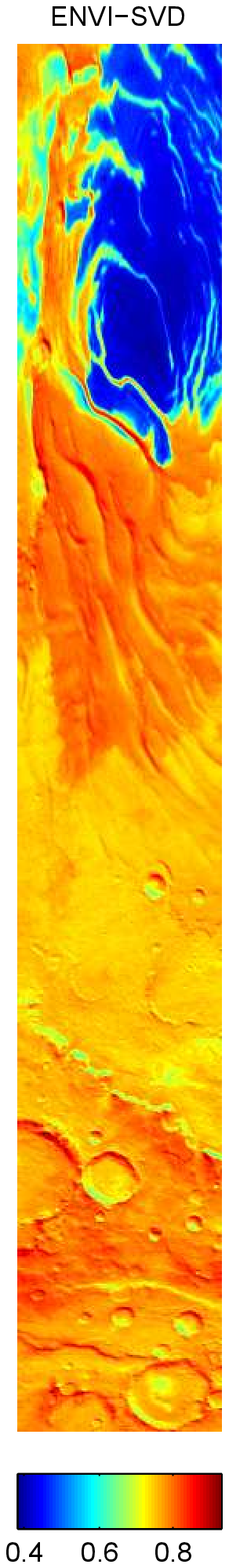}} \hspace{5mm}
\subfloat[]{\includegraphics[height=160mm]{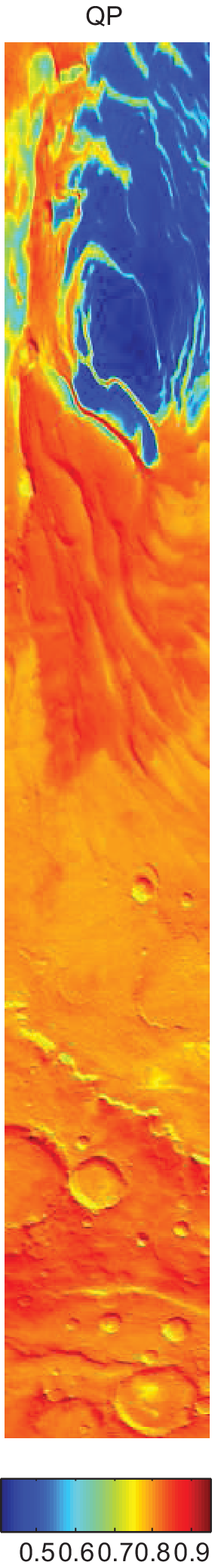}} \hspace{5mm}
\subfloat[]{\includegraphics[height=160mm]{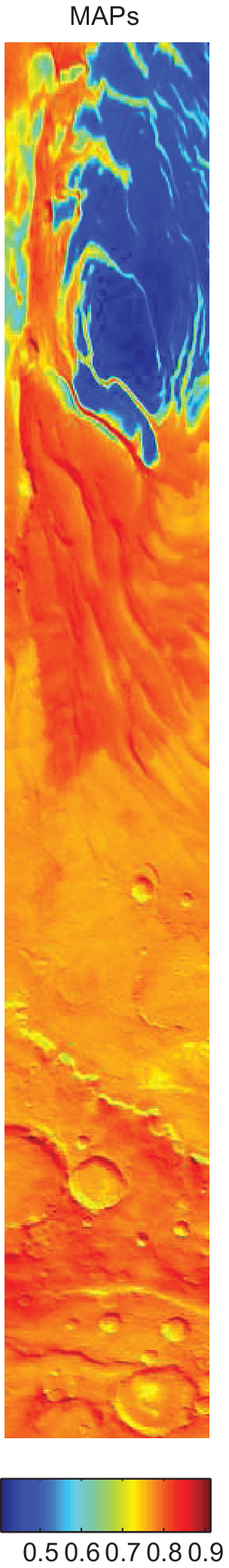}} \hspace{5mm}
\caption{Abundance map of the dust endmember, estimated using (a) ENVI-SVD, (b) QP, and (c) MAPs.}
\label{fig:endmember01}
\end{figure}
\begin{figure}[t]
\centering
\subfloat[]{\includegraphics[height=160mm]{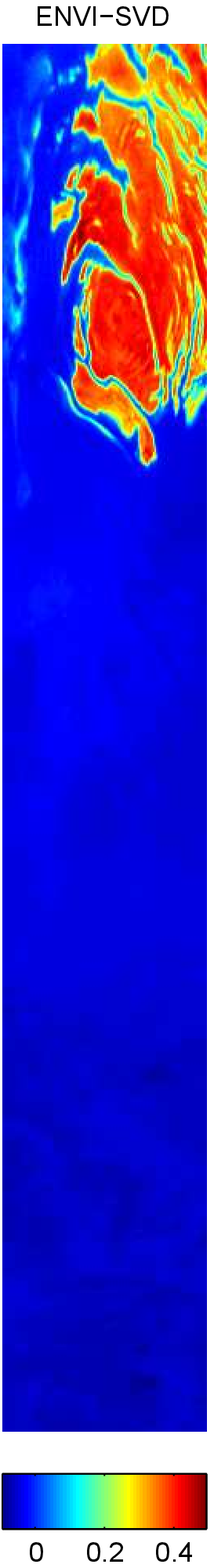}} \hspace{5mm}
\subfloat[]{\includegraphics[height=160mm]{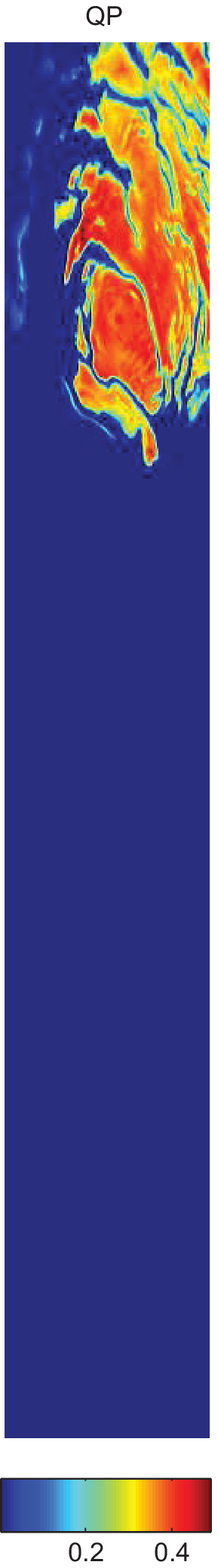}} \hspace{5mm}
\subfloat[]{\includegraphics[height=160mm]{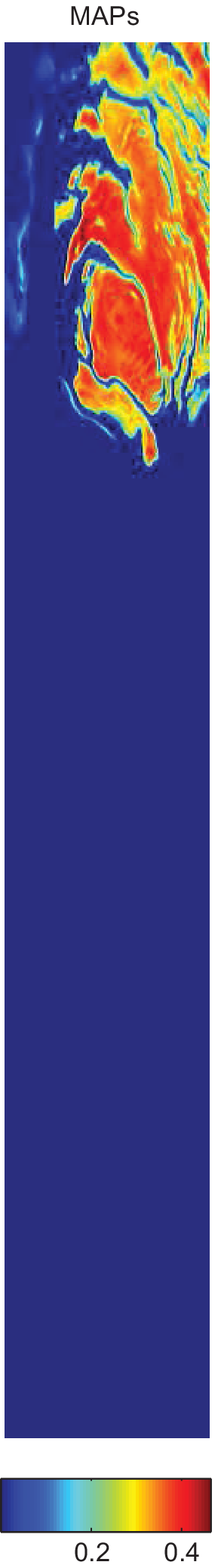}} \hspace{5mm}
\caption{Abundance map of the CO$_2$ endmember, estimated using (a) ENVI-SVD, (b) QP, and (c) MAPs.}
\label{fig:endmember02}
\end{figure}
\begin{figure}[t]
\centering
\subfloat[]{\includegraphics[height=160mm]{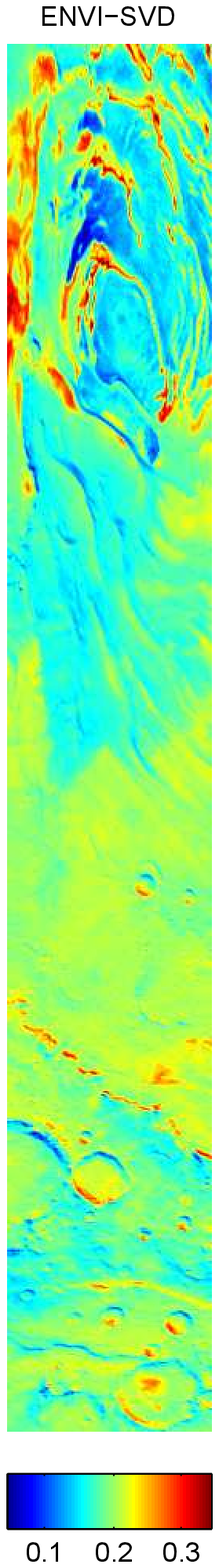}} \hspace{5mm}
\subfloat[]{\includegraphics[height=160mm]{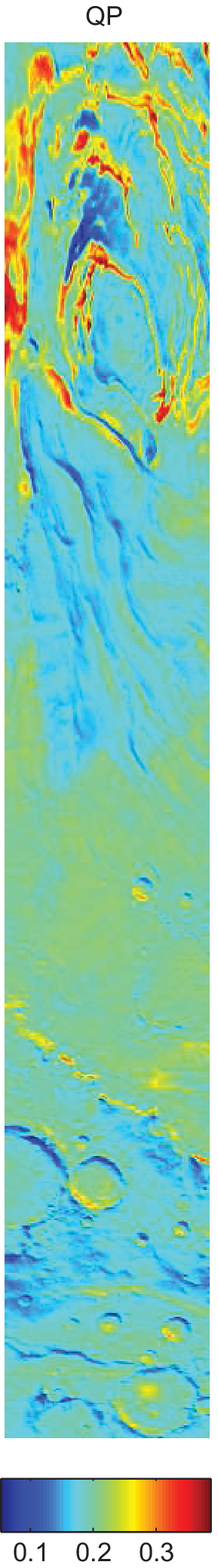}} \hspace{5mm}
\subfloat[]{\includegraphics[height=160mm]{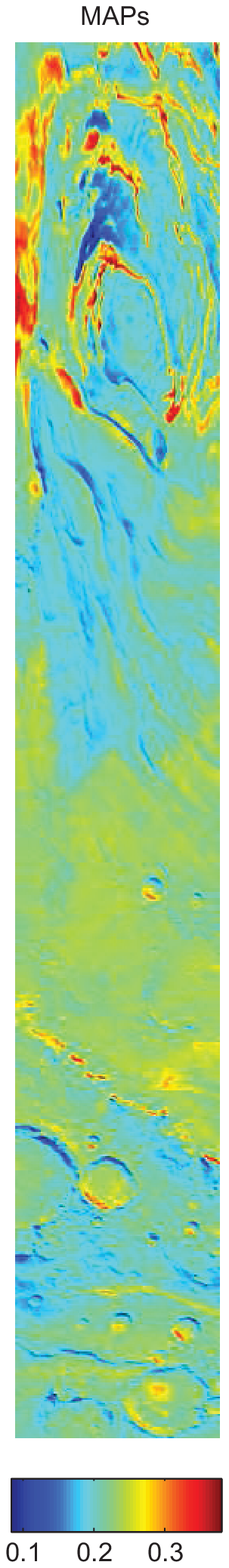}} \hspace{5mm}
\caption{Abundance map of the H$_2$O endmember, estimated using (a) ENVI-SVD, (b) QP, and (c) MAPs.}
\label{fig:endmember03}
\end{figure}

The abundance maps for each endmember resulting after the application of the three estimators are displayed in Figs. \ref{fig:endmember01} - \ref{fig:endmember03}. As shown in these figures, the ENVI-SVD abundances do not satisfy the non-negativity constraint, e.g., regarding the CO$_2$ endmember, the minimum computed abundance value is $-9.9 \times 10^{-2}$. Notice also that the abundance values calculated by QP and MAPs are in full agreement, they share the same scale and are quite different from those obtained by ENVI-SVD. This shows that the MAPs estimates provide reliable information about the abundances. It also adds up to the fact that the MAPs estimator has almost similar performance with the QP algorithm in simulation scenarios with synthetic data, as shown in \cite{2008_themelis}.



\subsection{Syrtis Major image cube}
\label{sec:syrtis}
\begin{figure}[t]
\centering
\includegraphics[width=\linewidth]{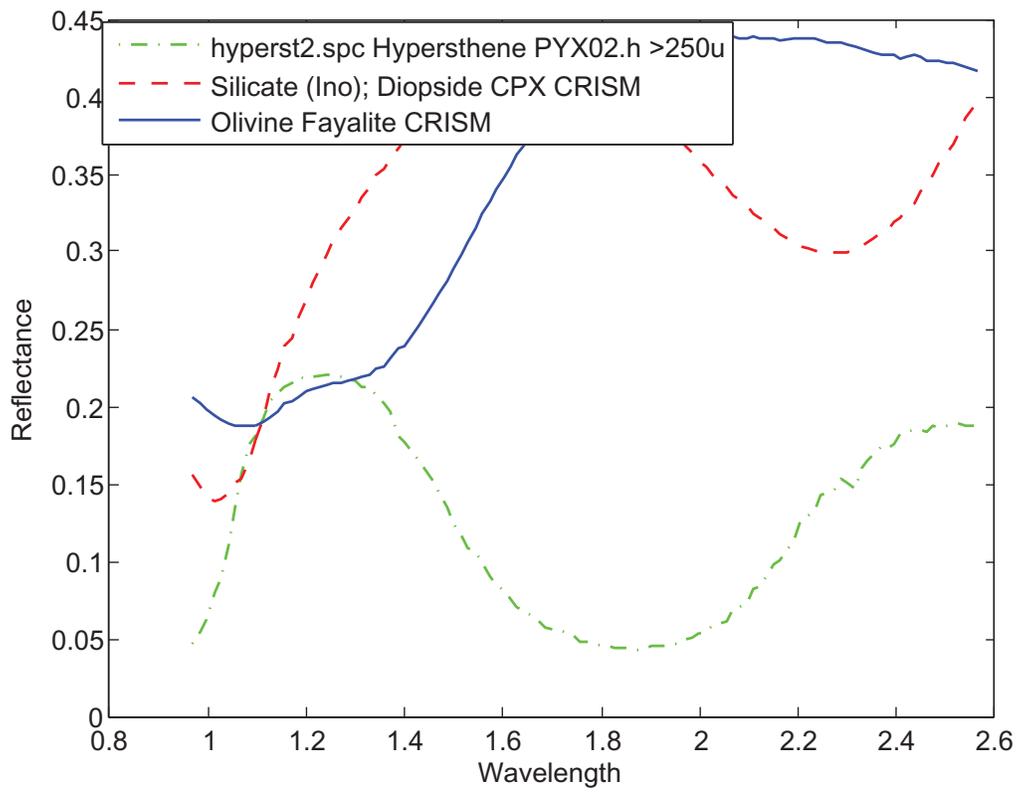}
\caption{Reference spectra of the Syrtis Major OMEGA image.}
\label{fig:endmbr_spectra02_real}
\end{figure}

This data set consists of a single hyperspectral data cube of the Syrtis Major region, which contains well-identified areas with very strong signatures of mafic minerals, \cite{2005_mustard}. The data cube consists of $109$ spectral bands out of the 128 original wavelengths of the SWIR detector. The spatial dimensions of the cube are $ 366 \times 128$ pixels. The OMEGA observations have been calibrated for known instrument artifacts and for atmospheric CO$_2$. The cube has been radiometrically corrected using the standard correction pipeline (SOFT06) and the atmospheric gas transmission has been empirically corrected using the volcano scan method, \cite{2005_langevin}. It is well known that OMEGA can identify pyroxene and olivine; it discriminates between the high-calcium pyroxenes (HCPs, e.g., clinopyroxenes) and low-calcium pyroxenes (LCPs, e.g., orthopyroxenes), \cite{2005_bibring}. In the scene under investigation, we utilized three endmembers which have previously been identified to be present in the image, \cite{2005_mustard}, namely, (a) Hypersthene, (b) Diopside (c) Fayalite. These are all laboratory reference spectra, which have also been used in \cite{2011_schmidt} for supervised unmixing. It is interesting to note that the last two endmembers have been retrieved using CRISM multispectral observations. Their respective spectral signatures are displayed in Fig. \ref{fig:endmbr_spectra02_real}. Three more artifact endmember spectra are utilized, specifically two neutral spectral components (flat lines at $10^{-4}$ and 1), and a slope line, as in \cite{2011_schmidt,2009_mouelic}.

\begin{figure}[t]
\centering
\subfloat[]{\includegraphics[height=120mm]{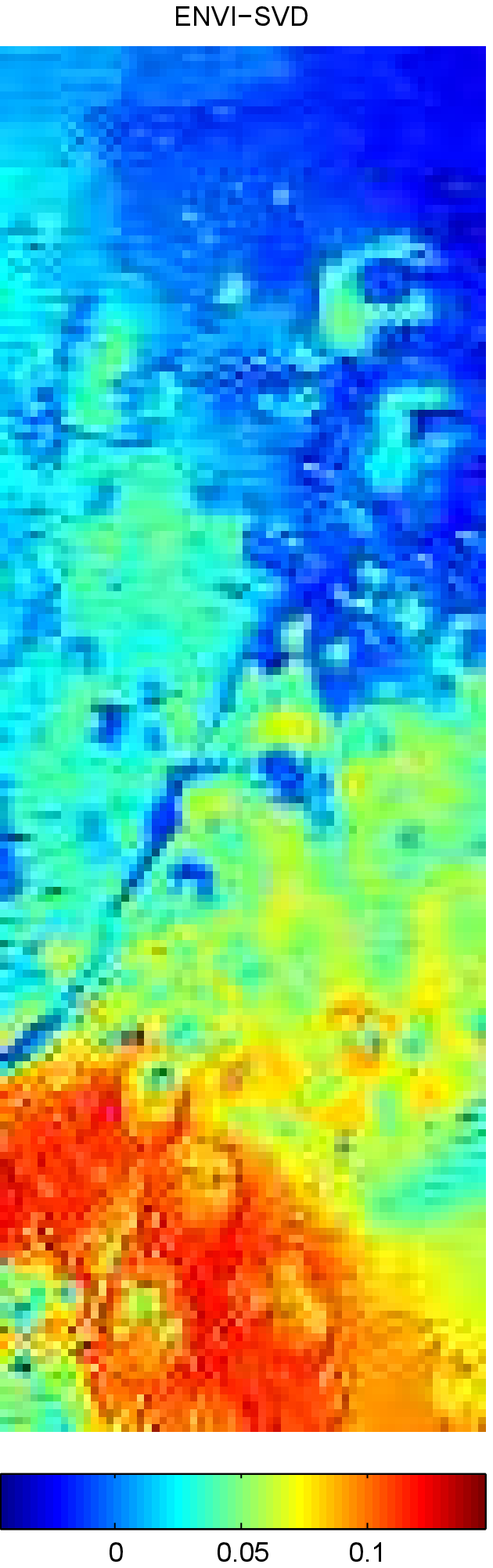}} \hspace{2mm}
\subfloat[]{\includegraphics[height=120mm]{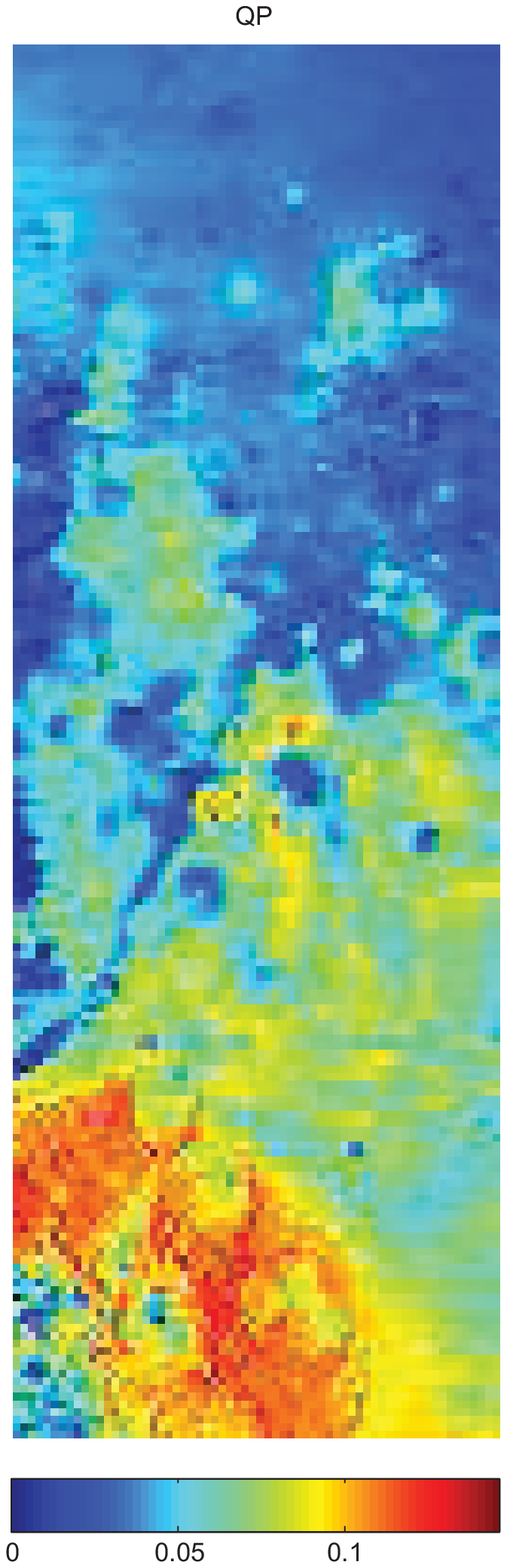}} \hspace{2mm}
\subfloat[]{\includegraphics[height=120mm]{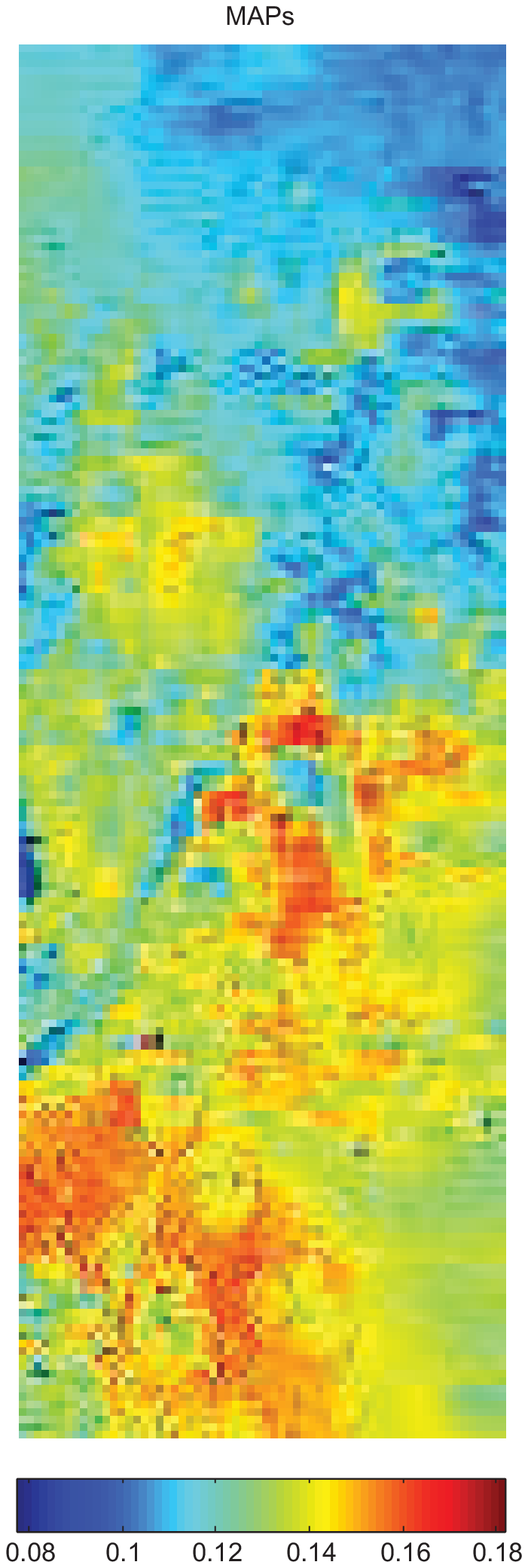}} \hspace{2mm}
\caption{Abundance map of Hypersthene, estimated using (a) ENVI-SVD, (b) QP, and (c) MAPs algorithms.}
\label{fig:endmember04}
\end{figure}

\begin{figure}[t]
\centering
\subfloat[]{\includegraphics[height=120mm]{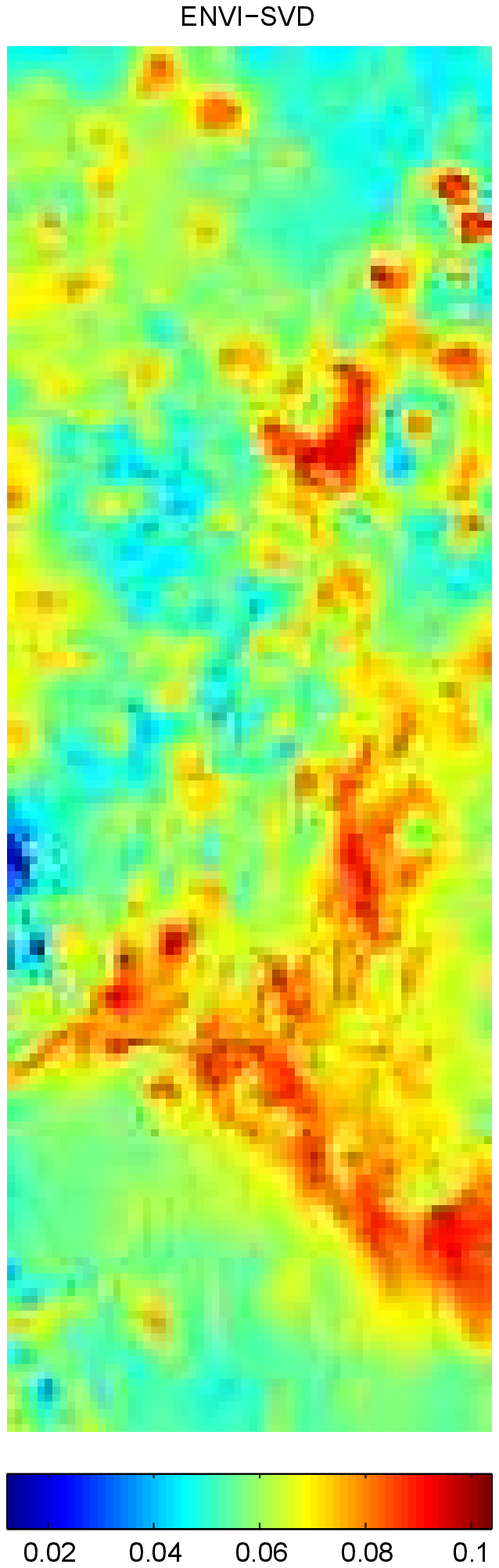}} \hspace{2mm}
\subfloat[]{\includegraphics[height=120mm]{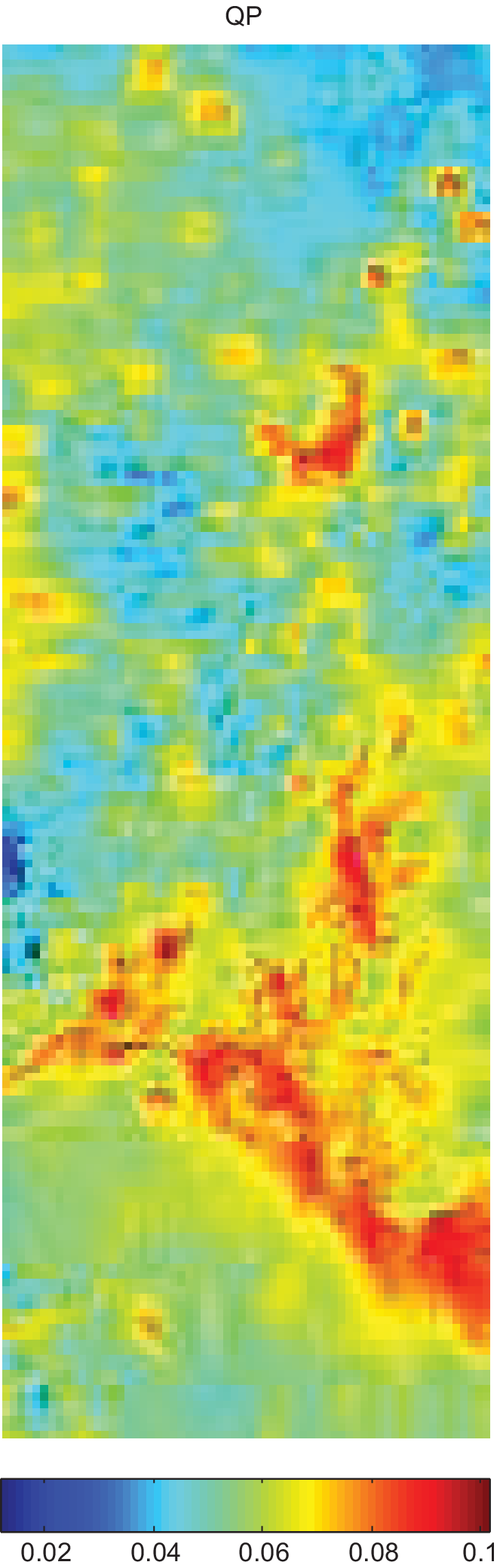}} \hspace{2mm}
\subfloat[]{\includegraphics[height=120mm]{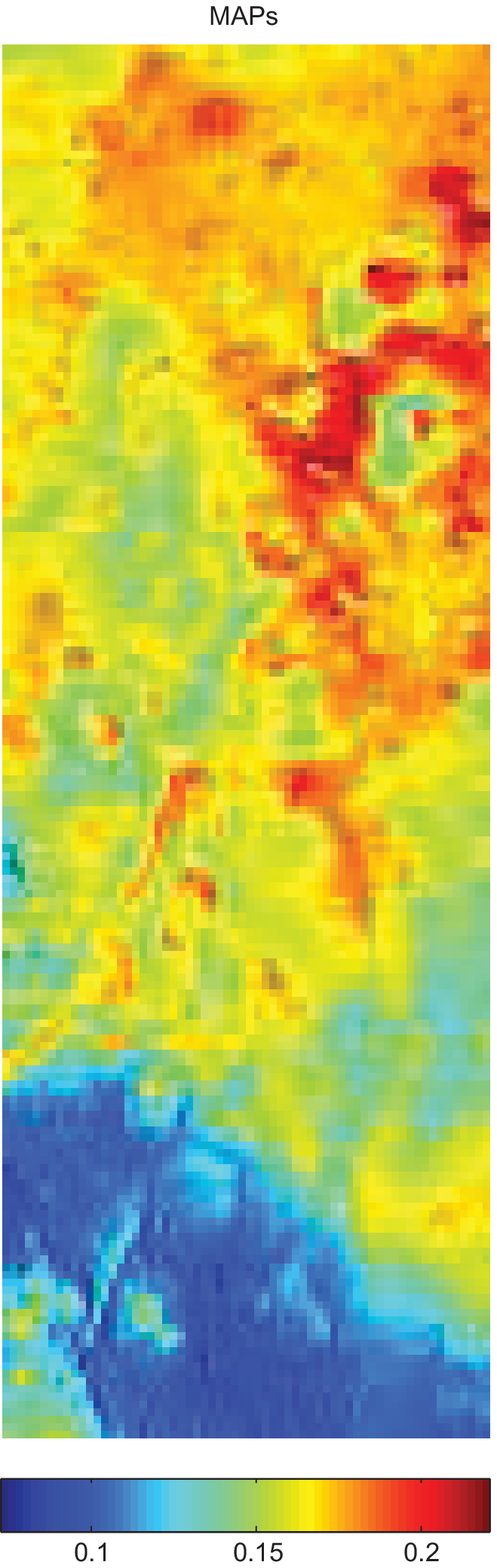}} \hspace{2mm}
\caption{Abundance map of Diopside, estimated using (a) ENVI-SVD, (b) QP, and (c) MAPs algorithms.}
\label{fig:endmember05}
\end{figure}

\begin{figure}[t]
\centering
\subfloat[]{\includegraphics[height=120mm]{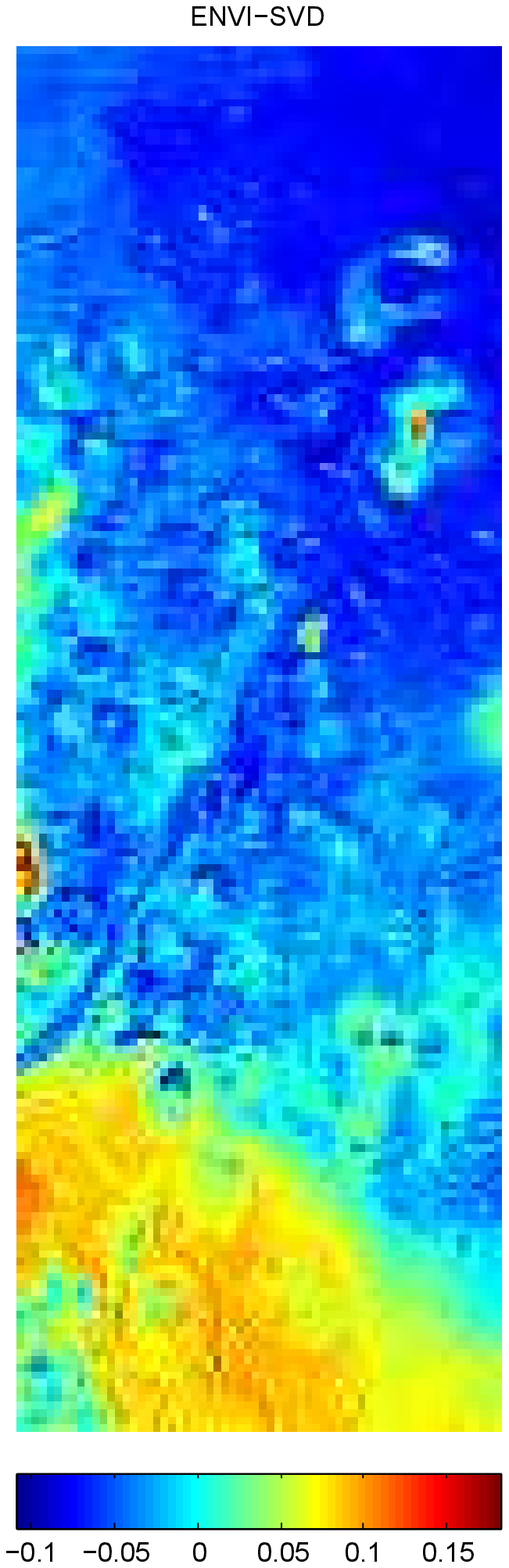}} \hspace{2mm}
\subfloat[]{\includegraphics[height=120mm]{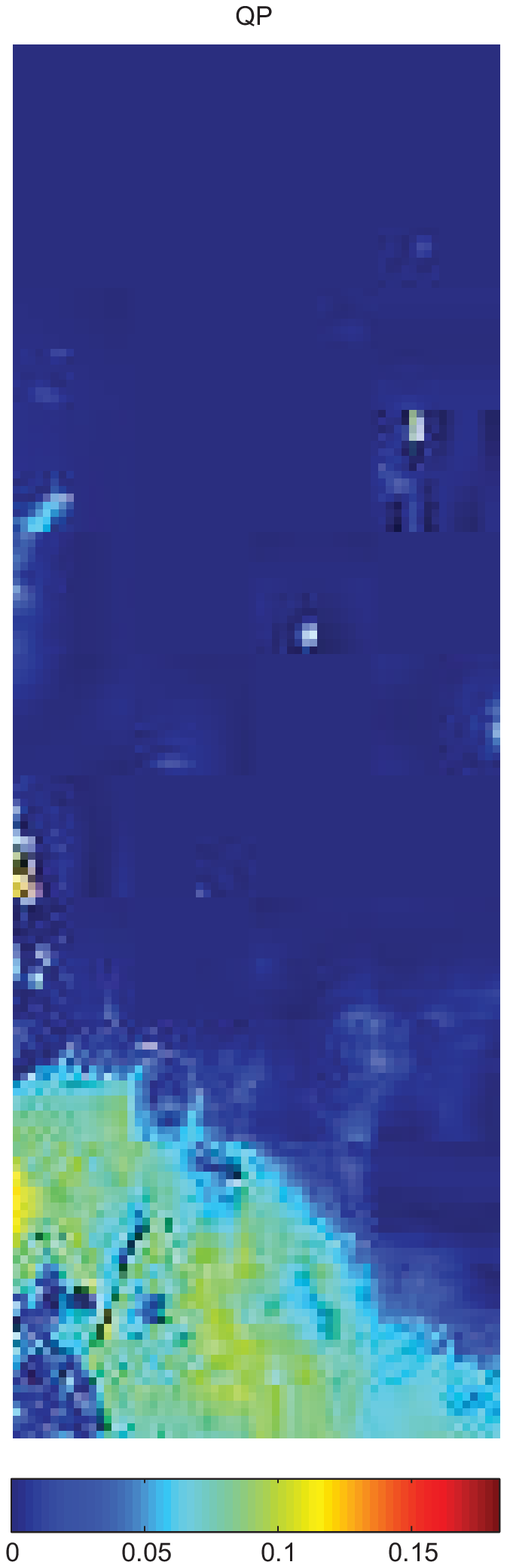}} \hspace{2mm}
\subfloat[]{\includegraphics[height=120mm]{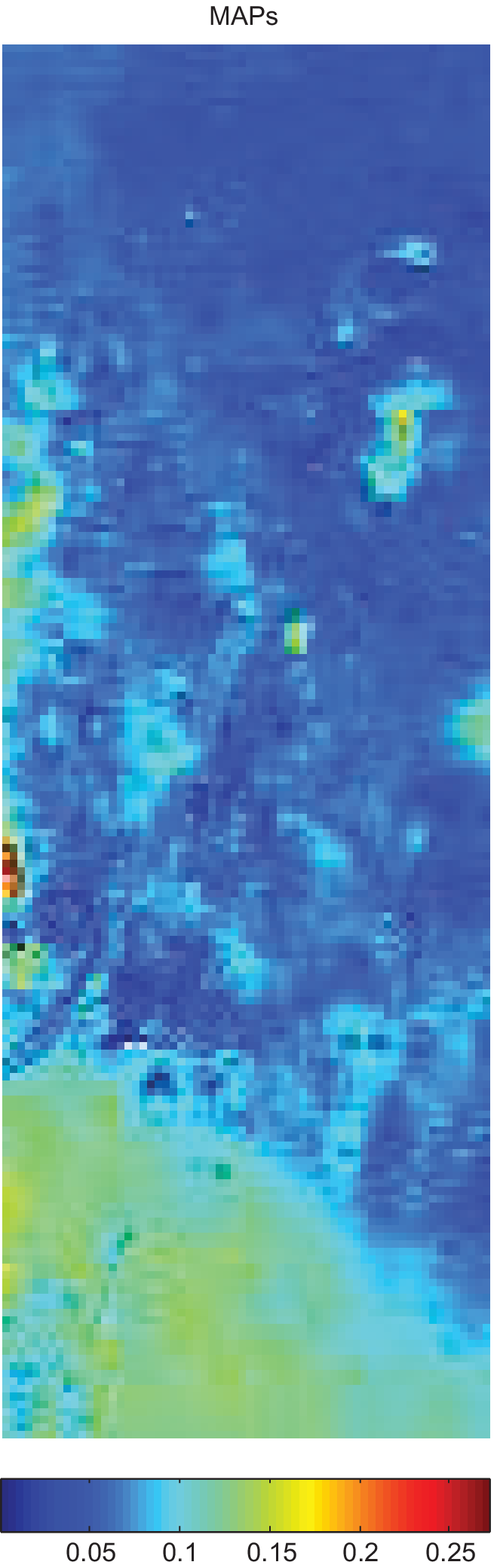}} \hspace{2mm}
\caption{Abundance map of Fayalite, estimated using (a) ENVI-SVD, (b) QP, and (c) MAPs algorithms.}
\label{fig:endmember06}
\end{figure}

\begin{figure}[t]
\centering
\subfloat[]{\includegraphics[height=120mm]{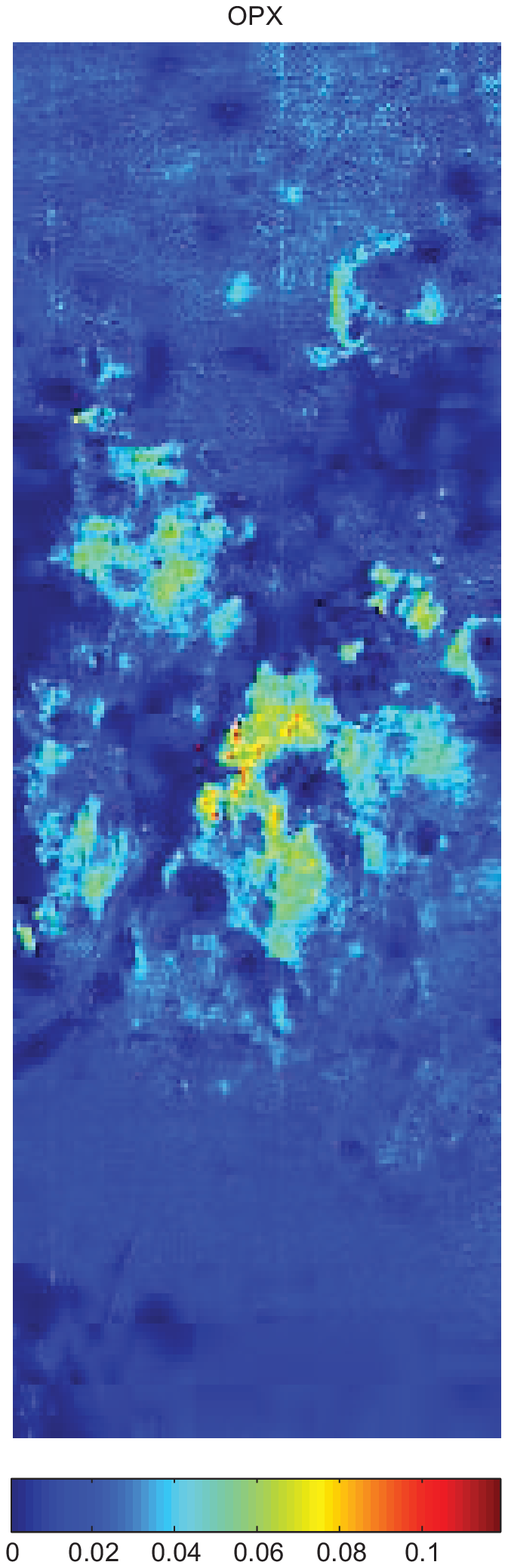}} \hspace{2mm}
\subfloat[]{\includegraphics[height=120mm]{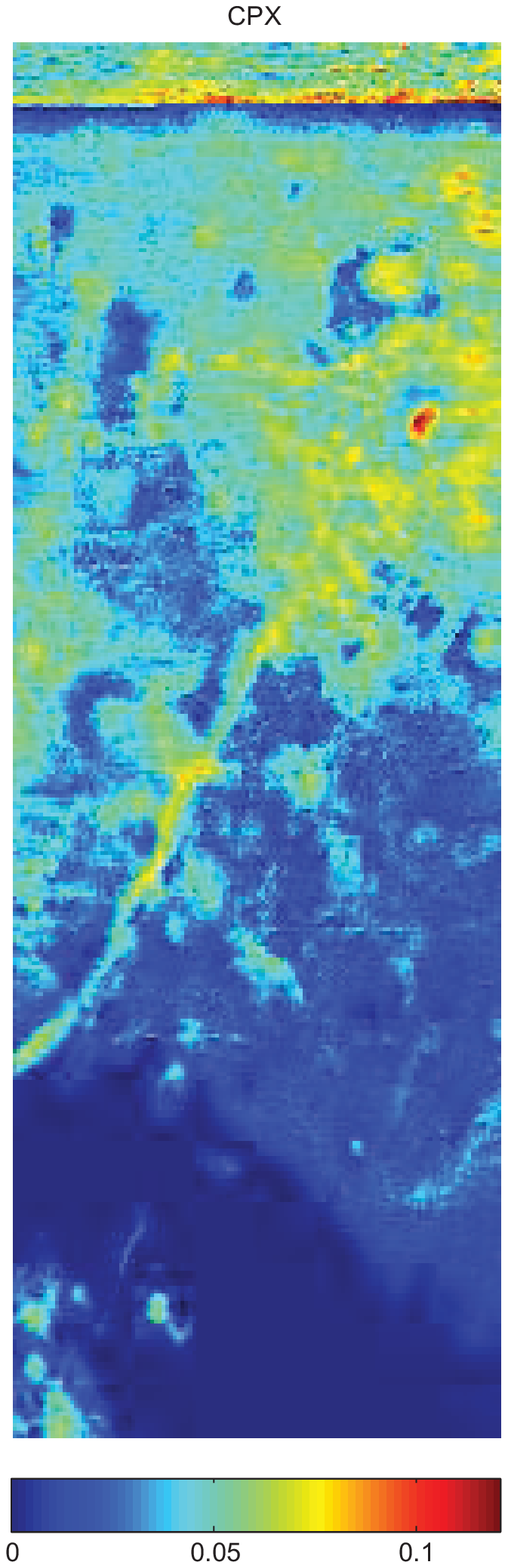}} \hspace{2mm}
\subfloat[]{\includegraphics[height=120mm]{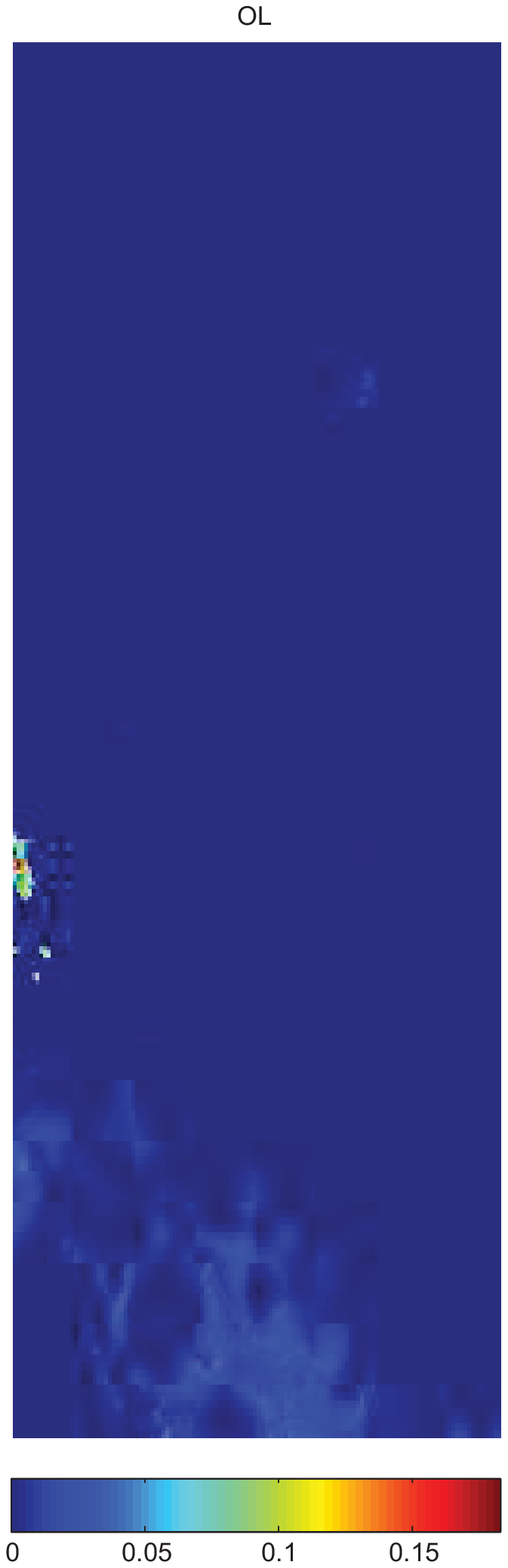}} \hspace{2mm}
\caption{Abundance maps of the endmembers (a) Hypersthene, (b)  Diopside, and (c) Fayalite in the Syrtis Major scene. The abundances are estimated using the band ratio method.}
\label{fig:endmember_ratio}
\end{figure}

The abundance maps obtained from the application of the three methods to the Syrtis Major hyperspectral scene are shown in Figs. \ref{fig:endmember04} - \ref{fig:endmember06}. Each Figure illustrates the corresponding abundance map of a single endmember, as it is estimated by all three methods. As a reference, the abundance maps of all three endmembers using the band ratio method are also shown in Fig. \ref{fig:endmember_ratio}. The band ratio and bands depth estimation methods are commonly used to detect minerals on OMEGA data \cite{2005_bibring}. This method is valid only if at least two wavelength channels can be identified as affected by only one particular mineral. Following the methodology of \cite{2011_schmidt}, we used four band ratios:
\begin{align}
{\rm Index(Olivine)}={\rm b}2.39/ {\rm b}1.06 
\end{align}
\begin{align}
 {\rm Index(opx)} &=1-{\rm b}1.84/((1.84-1.25)* {\rm b}1.25+(2.47-1.84)* \nonumber \\ 
& {\rm b}2. 47)*(2.47-1.25) 
\end{align}
\begin{align}
{\rm Index(cpx)}&=1-{\rm b}1.85/((2.32-1.85)*{\rm b}2.32+(2.56- 2.32) \nonumber \\ 
& *{\rm b}2.56)*(2.56-1.85)) 
\end{align}
where the wavelength band ``b$1.84$" stands for the band at $1.84$ microns, Orthopyroxenes (Hyperstene) are noted as ``opx" and Clinopyroxenes (Diopside) are noted as ``cpx". The differences between the three methods are again prominent. Both the MAPs and the QP methods provide consistent results, as far as the endmembers Fayalite and Hypersthene is concerned. In addition, for both Fayalite and Hypersthene, ENVI-SVD returns negative abundance values and its resulting maps substantially deviate from the other two methods. As far as Diopside is concerned, the MAPs estimator seems to produce a slightly different abundance map in comparison to the other two methods. However, it can be easily verified by comparing Figs. \ref{fig:endmember04} - \ref{fig:endmember06} and \ref{fig:endmember_ratio} that the abundance maps of the MAPs estimator are in better agreement with those obtained using the band ratio method, compared to the other two methods. Thus, it can be argued that MAPs provides more reliable results than ENVI-SVD and QP.





\section{Conclusions}
\label{sec:disc}
In this paper, we have presented a comparison of three different supervised spectral unmixing methods (Bayesian MAPs estimator, ENVI-SVD, QP), on the basis of two different OMEGA hyperspectral data sets. 
As opposed to iterative algorithms or Markov Chain Monte Carlo methods, e.g. \cite{2010_schmidt}, commonly used for the constrained inverse problem of abundance estimation, the computational complexity of the MAPs estimator is much lower. Specifically, for the Syrtis Major dataset, the running time of the MAPs algorithm was $4.7$ secs in a roughly optimized Matlab implementation, while the QP needed 26.4 secs (both algorithms were run on a $2.4$-Ghz Intel Core $2$ CPU). In addition, as verified by experimental results, the performance of the two methods is approximately equal. Therefore, the MAPs estimator seems to offer the best compromise between estimation performance and complexity among the three algorithms, and it is thus a serious candidate for spectral unmixing of hyperspectral data in Planetary Sciences. 




\section{Acknowledgment}
The authors would like to thank ESA and the OMEGA team for providing us the hyperspectral data used in this paper.

\bibliographystyle{elsarticle-harv}
\bibliography{bibl}







\end{document}